\documentclass[runningheads]{llncs}
\usepackage[T1]{fontenc}
\usepackage{graphicx}
\usepackage{amsmath}
\usepackage{amssymb}
\usepackage{amsfonts}
\usepackage{mathtools}
\usepackage{subcaption}
\usepackage{booktabs}
\usepackage{adjustbox}
\usepackage{multirow}
\usepackage{xcolor, colortbl}
\usepackage{hyperref}
\usepackage{algorithm}
\usepackage{algpseudocode} 

\usepackage{amsmath}

\definecolor{cgreen}{rgb}{0,0.7,0.8}
\definecolor{cred}{rgb}{0.968,0.545,0.321}

\newcommand{\rtr}[1]{{\scriptsize\color{cred}$\blacktriangledown$ #1}}

\newcommand{\gtr}[1]{{\scriptsize\color{cgreen}$\blacktriangle$ #1}}
\begin{document}
\title{Diffusion Models with Implicit Guidance for Medical Anomaly Detection}
\titlerunning{THOR}
%
%
%
\author{Cosmin I. Bercea\inst{1,2} \and
Benedikt Wiestler\inst{3} \and
Daniel Rueckert\inst{1,3,5}
\and
Julia A. Schnabel\inst{1,2,4}}
\authorrunning{CI. Bercea et al.}
\institute{Technical University of Munich, Munich, Germany \and Helmholtz AI and Helmholtz Center Munch, Munich, Germany \and Klinikum Rechts der Isar, Munich, Germany \and Kings College London, London, UK \and Imperial College London, London, UK
}
\maketitle              
\begin{abstract}

Diffusion models have advanced unsupervised anomaly detection by improving the transformation of pathological images into pseudo-healthy equivalents. Nonetheless, standard approaches may compromise critical information during pathology removal, leading to restorations that do not align with unaffected regions in the original scans. Such discrepancies can inadvertently increase false positive rates and reduce specificity, complicating radiological evaluations. This paper introduces Temporal Harmonization for Optimal Restoration (\textit{THOR}), which refines the de-noising process by integrating implicit guidance through temporal anomaly maps. \textit{THOR} aims to preserve the integrity of healthy tissue in areas unaffected by pathology. 
Comparative evaluations show that \textit{THOR} surpasses existing diffusion-based methods in detecting and segmenting anomalies in brain MRIs and wrist X-rays. Code: \url{https://github.com/ci-ber/THOR_DDPM}.

\begin{figure}[b!]
    \centering
    \includegraphics[width=0.929\textwidth]{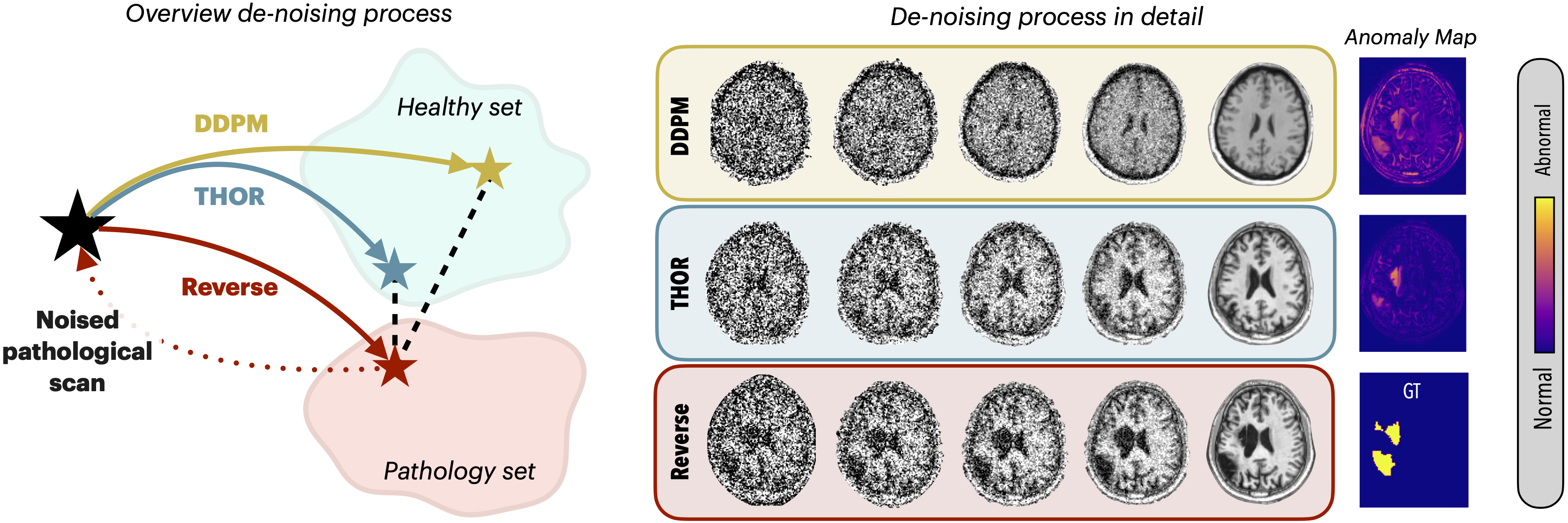} 
    \caption{Denoising diffusion probabilistic models (DDPMs) trend towards a generalized healthy reference. From a noised starting point, \textit{THOR} employs temporal harmonization to yield an outcome resembling the original in healthy tissue regions, concurrently replacing pathology areas with pseudo-healthy restorations.}
    \label{fig::thor_teaser}
\end{figure}

\end{abstract}
\section{Introduction}
Robust and accurate anomaly detection is vital for early diagnosis and effective treatment, especially in the face of rare and diverse pathologies. The complexity and variability inherent in medical conditions present substantial challenges to conventional diagnostic methods anchored in supervised learning~\cite{kamnitsas2016deepmedic,zhou2023foundation}. These methods depend heavily on extensive, annotated datasets, which are difficult to obtain for rare conditions, limiting the scope and flexibility of diagnostic tools. In response, unsupervised learning has emerged as a viable alternative, capable of detecting anomalies across a broad spectrum without the need for explicit labels~\cite{bercea2022ra,zimmerer2019unsupervised,pinaya2022unsupervised,tan2021detecting}. Among unsupervised techniques, denoising diffusion probabilistic models (DDPMs)~\cite{ho2020denoising} have shown substantial promise in enhancing the precision and efficiency of anomaly detection. By adding and subsequently removing noise, DDPMs transform pathological inputs into pseudo-healthy outputs, demonstrating impressive generative potential. Nonetheless, this noise-dependent process can result in significant loss of information, leading restored images to deviate from their original state, including in regions unaffected by pathology. Such deviations risk increasing false positives and decreasing specificity, further complicating the diagnostic process.

To overcome the limitations inherent in DDPMs, more sophisticated models have been developed. AnoDDPM proposes to use Simplex noise, which allows the use of lower noise levels~\cite{Wyatt_2022_CVPR}. Conditional diffusion models blend the capabilities of autoencoders with diffusion techniques to incorporating semantic information such as tissue intensity into the de-noising process~\cite{behrendt2023guided}. Patch-based DDPMs (pDDPMs) extend these advancements by applying the diffusion process to localized patches of the image, using adjacent areas as contextual anchors in a sliding-window technique~\cite{behrendt2023patched}. AutoDDPMs build upon this foundation with a unique approach that involves masking, stitching, and re-sampling, utilizing dual de-noising processes at different levels of noise to seamlessly integrate context into the reconstructions~\cite{bercea2023mask}. While these innovations represent substantial progress, they also introduce complexities. The task of determining an optimal patch size that can adapt to the multiple scales of diseases is challenging due to the diversity of pathological presentations. Additionally, the complexity of orchestrating dual de-noising processes across different noise levels requires precise calibration. These challenges could potentially limit their practicability.

Diffusion models enhanced with classifier guidance use weakly supervised classifiers for anomaly detection, leveraging gradients to refine the identification of anomalous regions~\cite{wolleb2022diffusion}. However, the effectiveness of this approach depends on the accuracy of classifiers, potentially limiting its capability to detect diseases independently by biasing it towards known pathologies.

In this work, we introduce \textit{THOR} (Temporal Harmonization for Optimal Restoration), a novel approach designed to enhance unsupervised anomaly detection in medical imaging, as illustrated in Fig.~\ref{fig::thor_teaser}. \textit{THOR} incorporates implicit guidance into diffusion models through the use of temporal anomaly maps, aiming to preserve the original image context while achieving accurate anomaly detection and segmentation. Our key contributions are as follows:
\begin{itemize}
    \item[$\bullet$] We develop \textit{THOR} and leverage implicit guidance within diffusion models to facilitate optimal image restorations and improve the accuracy of anomaly segmentation.
    \item[$\bullet$] We apply \textit{THOR} to two challenging medical datasets, where it demonstrates its capability in accurately segmenting stroke lesions and localizing pathology in pediatric wrist X-rays, thereby enhancing performance in essential diagnostic tasks.
    \item[$\bullet$] We perform a sensitivity analysis of critical hyperparameters such as different noise types and levels.
\end{itemize}

\section{Background}

\subsection{Anomaly Detection Setup}

In medical imaging anomaly detection, the objective is to detect deviations from normal anatomical structures without explicit pathological labels. We define $\tilde{X}$ as the domain of all medical images, where each image $\tilde{x} \in \tilde{X}$ includes regions of both normal and abnormal tissue. The aim is to assign an anomaly score $S$ to each pixel (or voxel), using a function $f: \tilde{X} \rightarrow S$.

Considering a dataset of medical images ${x}_{i=1}^{N} \in X \subset \tilde{X}$ for training, these images are presumed to represent a healthy tissue distribution, denoted as $P(X)$. The challenge lies in accurately modeling $P(X)$. By doing so, we can project any input image into the $P(X)$ space, creating a pseudo-healthy reconstruction. If an input image has pathology, this method produces a version where pathological features are replaced with those typical of healthy tissue according to $P(X)$. This approach enables anomaly detection by contrasting the original image with its pseudo-healthy counterpart to identify deviations.

\subsection{Denoising Diffusion Probabilistic Models (DDPMs)}

DDPMs are generative models that aim to replicate the distribution $P(X)$ through a process that incrementally introduces and reverses Gaussian noise.

\textbf{Forward Process.} In the forward process, a DDPM gradually transforms a clean image $x_0$, drawn from the distribution $P(X)$, into a completely noisy state over a Markov chain of $T$ steps, described by:
\begin{equation}
x_t = \sqrt{\alpha_t} x_{t-1} + \sqrt{1 - \alpha_t} \epsilon, \quad \epsilon \sim \mathcal{N}(0, \mathbf{I}),
\label{eq::noise_forward}
\end{equation}
where $\alpha_t$ is part of a predetermined noise schedule (0 < $\alpha_t$ < 1), and $\epsilon$ represents Gaussian noise. The sequence ${x}_{t=0}^{T}$ depicts the transition of the input image into a state where $x_T$ is predominantly noise.

\textbf{Reverse Process.} The reverse process aims to reconstruct the clean image from its noisy counterpart by denoising, essentially learning $P(X)$. This can be formulated as:
\begin{equation}
x_{t-1} = \frac{1}{\sqrt{\alpha_t}} \left(x_t - \sqrt{1 - \alpha_t} \epsilon_{\theta}(x_t, t) \right),
\label{eq::noise_reverse}
\end{equation}
where $\epsilon_{\theta}(x_t, t)$ is the estimate of the noise added at step $t$. By learning $\epsilon_{\theta}$, the model inverts the noising process, approximating the distribution of $P(X)$.
\section{Method: THOR}
\begin{figure}[tb!]
    \centering
    \includegraphics[width=0.999\textwidth]{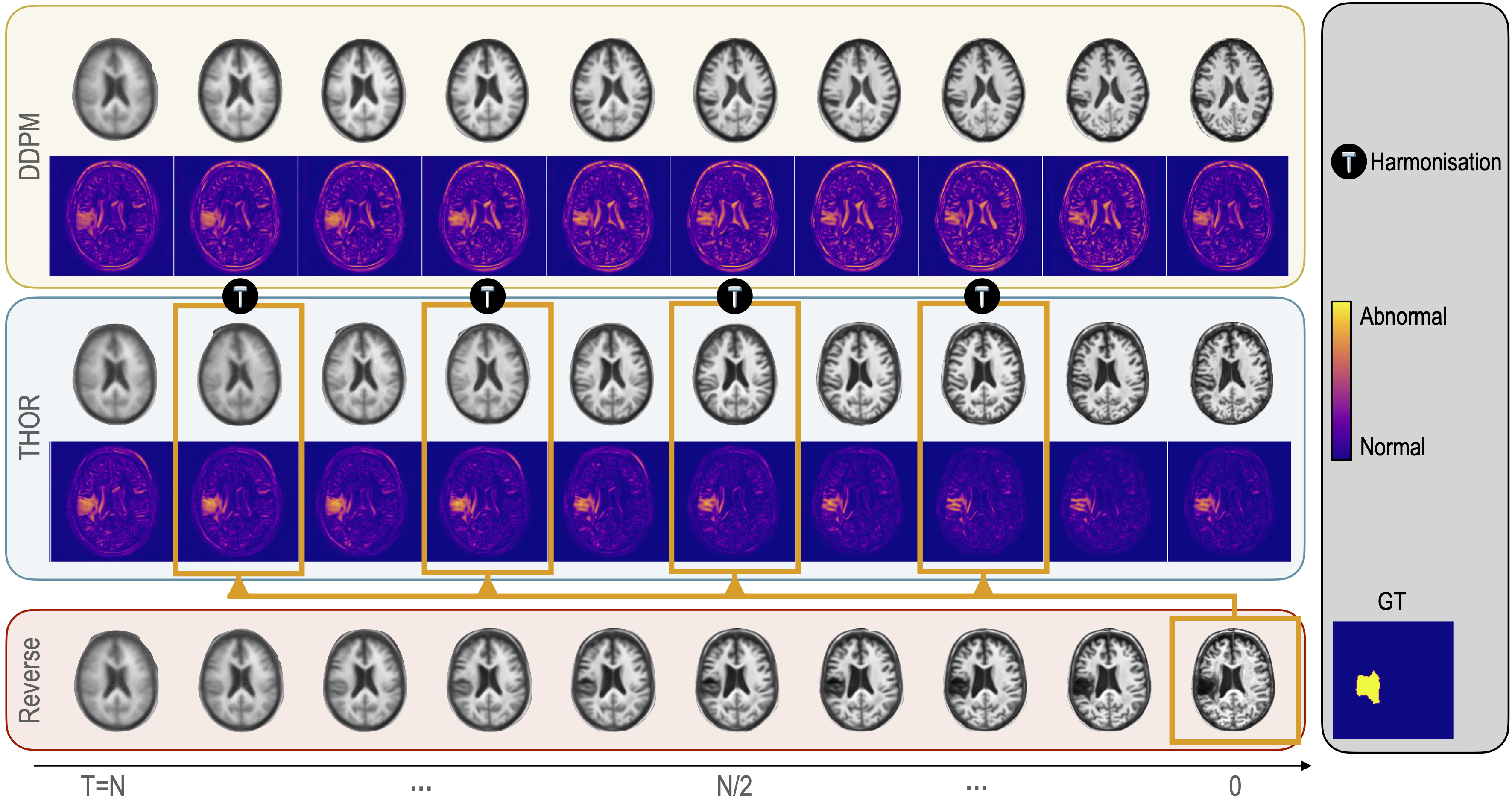} 
    \caption{The top row displays the traditional DDPM de-noising sequence, where noise is progressively reduced to clarify image features. In contrast, the middle row showcases \textit{THOR}: starting with equivalent high noise levels and then strategically applying unsupervised temporal anomaly masks at key intervals (indicated by orange borders) to 'harmonize' the image. This 'harmonization' process selectively refines the image by maintaining normal tissue integrity while attenuating anomalies. The bottom row shows the reverse process with anomalies becoming increasingly apparent as noise is reversed, culminating in the ground truth (GT) image where the anomaly is clearly delineated.}
    \label{fig::thor_method}
\end{figure}
\textit{THOR} advances the de-noising process in DDPMs, offering guidance during inference through the application of unsupervised temporal anomaly masks, without necessitating retraining. Typically, DDPMs necessitate high noise levels ($T$) to effectively obscure anomalies, a practice that can compromise the integrity of non-pathological tissue details. Such an approach may result in the loss of critical anatomical information, thereby elevating the potential for false positives. The innovation of \textit{THOR} lies in its ability to guide the restoration process by strategically reintegrating healthy tissue information, a technique we refer to as "harmonization." This method starts at the same elevated noise levels but diverges by using implicit temporal anomaly masks to inform the denoising trajectory. Such guidance aims to selectively restore the image, focusing on preserving the fidelity of non-pathological regions while reducing the anomalies. Details of our procedural approach are delineated in Fig.~\ref{fig::thor_method}.\\ 

\noindent\textbf{Implicit Guidance via Intermediate Anomaly Maps.}
Intermediate anomaly maps play an essential role in the unsupervised "harmonization" process of \textit{THOR}, applied at specific timesteps. These maps critically compare the predictive reconstructions \( x_0^t \) with the actual input image \( x_0^{\text{input}} \), highlighting discrepancies that indicate anomalies and distinguishing regions that are likely healthy. Anomaly maps \( m \) combine residual differences with the Learned Perceptual Image Patch Similarity (LPIPS) metric~\cite{zhang2018unreasonable}, enhancing the identification of subtle pathological changes~\cite{bercea2022ra}:
\begin{equation}
m(x,x_{rec}) = \lvert x - x_{rec} \rvert \cdot S_{\text{LPIPS}}(x,x_{rec}).
\label{eq::anomaly_map}
\end{equation}

To avoid incorporating anomalous regions in the denoising process, we normalize the values of \( m \) between 0 and 1 and apply morphological operations, specifically a sequence of closing followed by dilation (denoted as \( cd \)).

These anomaly maps are then utilized in the "harmonization" process to adjust the interpolation between the pseudo-healthy predictions and the actual inputs. This adjustment aims to producing reconstructions that not only closely resemble the original images but also conform to the healthy tissue profile:
\begin{equation}
x_{t} = cd(m(x_0^t, x_0^{\text{input}})) \cdot x_0^{\text{prediction}} + (1 - cd(m(x_0^t, x_0^{\text{input}}))) \cdot x_0^{\text{input}}.
\label{eq::interpolation}
\end{equation}

The final anomaly score, $S$, is calculated using the harmonic mean of the anomaly maps at the selected timesteps, explicitly defined as: 
\begin{equation}
    S = n \bigg/ \sum_{t \in \text{{selected steps}}} \frac{1}{m(x_0^t, x_0^{\text{{input}}})},
\label{eq::anomaly_score}
\end{equation}
where $n$ is the total number of selected harmonization timesteps.


\section{Experiments and Results}
To demonstrate the utility, generalizability, and performance of our method, we conduct two experiments, i.e., the segmentation of ischemic stroke lesion in MRI in Sec.~\ref{sec::stroke} and anomaly localization in pediatric wrist X-ray images in Sec.~\ref{sec::wxr}.
\subsection{Ischemic Stroke Lesion Segmentation in Brain MRI \label{sec::stroke}}

Stroke represents a major cause of disability and mortality worldwide, with its early detection being paramount for effective treatment planning. 


\noindent\textbf{Datasets}. The training dataset encompasses 582 T1-weighted MRI scans from the IXI~\cite{ixi} dataset and 217 healthy samples from the ATLAS v2.0 dataset~\cite{atlas2022}, offering a wide representation of normal brain anatomy. For testing, we employed the ATLAS dataset, which includes 655 T1w MRI scans with expertly segmented lesion masks. We categorized the anomalies into small (less than 71 pixels), medium, and large ($\geq 570$) lesions for detailed analysis, excluding 20 slices with significant unannotated hypo-intense artifacts to maintain data integrity. Scans were normalized to the 98th percentile and resized to $128\times128$ pixels, with lesion segmentation evaluated via the maximum achievable Dice. \\

\begin{table}[t!]
    
    \centering
    \setlength{\tabcolsep}{4pt}
    \caption{\textbf{Performance on Brain MRI Stroke Segmentation.} \textit{THOR}, our proposed method, considerably outperforms other methods (DDPM, AutoDDPM, AnoDDPM, pDDPM) across different lesion sizes, marked by the {\boldmath $bold$} numbers and percentage improvements (\gtr{x}) compared to the best baseline. \label{tab::stroke}}
    \begin{adjustbox}{width=0.85\linewidth,center} 
        \begin{tabular}{l l | c || c c c}
            \toprule	    
           \multirow{2}{*}{\rotatebox{90}{Noise}} & \multirow{2}{*}{Method} & \multicolumn{4}{c}{Pathology $\lceil Dice \rceil$ $\uparrow$}\\
           & & Average & Small & Medium  & Large\\\midrule
            \multirow{3}{*}{\rotatebox{90}{Gauss}} & \cellcolor{gray!10} THOR (ours) & \cellcolor{gray!10} {\boldmath$20.41$}~\gtr{20\%} & \cellcolor{gray!10} {\boldmath$9.14$}~\gtr{103\%} &  \cellcolor{gray!10} {\boldmath$26.34$}~\gtr{19\%} & \cellcolor{gray!10} $41.26$~\rtr{5\%}   \\ 
     	   & DDPM~\cite{ho2020denoising}  & 8.05~\rtr{61\%} & $1.37$~\rtr{85\%} & $9.53$~\rtr{64\%}  & $25.65$~\rtr{38\%} \\ 
     	   & AutoDDPM~\cite{bercea2023mask} & 16.95~\rtr{17\%} & $4.55$~\rtr{50\%} & $22.07$~\rtr{16\%}  & \boldmath{$43.47$}~\gtr{5\%} \\ \midrule
            \multirow{3}{*}{\rotatebox{90}{Simplex}} &  \cellcolor{gray!10} THOR (ours) & \cellcolor{gray!10} {\boldmath $29.74$}~\gtr{33\%} & \cellcolor{gray!10} {\boldmath$11.54$}~\gtr{44\%}  & \cellcolor{gray!10} {\boldmath$39.20$}~\gtr{30\%}  & \cellcolor{gray!10} {\boldmath $63.64$}~\gtr{34\%} \\
             
     	   & AnoDDPM~\cite{Wyatt_2022_CVPR}  & 18.07~\rtr{39\%} & 4.82~\rtr{58\%} & 23.45~\rtr{40\%} & 46.65~\rtr{27\%} \\ 
      	  & pDDPM~\cite{behrendt2023patched}  &  22.28~\rtr{25\%} & {$8.02$}~\rtr{31\%} & $30.16$~\rtr{23\%}  & $47.66$~\rtr{25\%} \\ 
     	    \bottomrule
\end{tabular}
    \end{adjustbox}
\end{table}
        
\begin{figure}[tb!]
    \centering
    \includegraphics[width=0.999\textwidth]{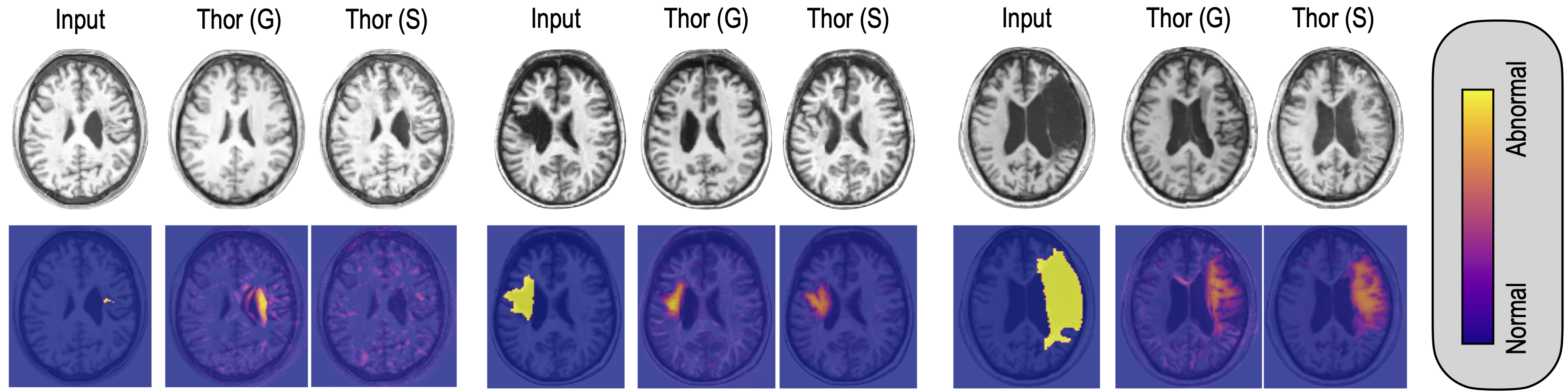} 
    \caption{Anomaly detection in brain MRI scans processed by \textit{THOR} using Gaussian (G) and Simplex (S) noise. From left to right, the lesions increase in size, with the smallest representing a challenging case.
}
    \label{fig::qualitative}
\end{figure}

\noindent\textbf{Results.} Tab.~\ref{tab::stroke} shows quantitative results and explores two key diffusion noise scenarios: Gaussian and Simplex. This examination is vital for assessing the performance of \textit{THOR} in comparison with leading diffusion models. \textit{THOR} is proficient with both types of noise, illustrating its broad applicability.

\textit{Gaussian noise} is the conventional choice for DDPMs but introduces challenges in anomaly detection. 
Due to the partial denoising strategy employed for anomaly detection, a high noise level (here T=350) is essential to effectively conceal anomalies~\cite{Wyatt_2022_CVPR}. Yet, deploying Gaussian noise at such high iterations frequently results in false positives due to inaccuracies in restoring healthy tissue. This limitation is reflected in the diminished segmentation scores for DDPM. Conversely, our harmonization process navigates the de-noising towards more precise restorations. Consequently, \textit{THOR} addresses the challenge of false positives and significantly refines the accuracy of anomaly segmentation, as evidenced both numerically in Tab.~\ref{tab::stroke} and visually in Fig.~\ref{fig::thor_method} and Fig.~\ref{fig::qualitative}.

\textit{Simplex noise} provides a notable advantage with its coarse noise patterns, allowing for the de-noising process to commence at lower levels (T=250) as demonstrated in~\cite{Wyatt_2022_CVPR}. This characteristic is beneficial, preserving more of the original image context and laying a stronger groundwork for restoration. The utility of Simplex noise becomes apparent when observing the improved performance of models like AnoDDPM, which exhibit significant enhancements over the traditional DDPM. Leveraging the capability of Simplex noise, \textit{THOR} advances the restoration process further. Its harmonization process meticulously refines the output, ensuring restorations more faithfully represent the original healthy tissue and thereby outperforming AnoDDPM and similar models (see Tab.~\ref{tab::stroke}).\\

\begin{figure}[t!]
    \centering
    \includegraphics[width=0.999\textwidth]{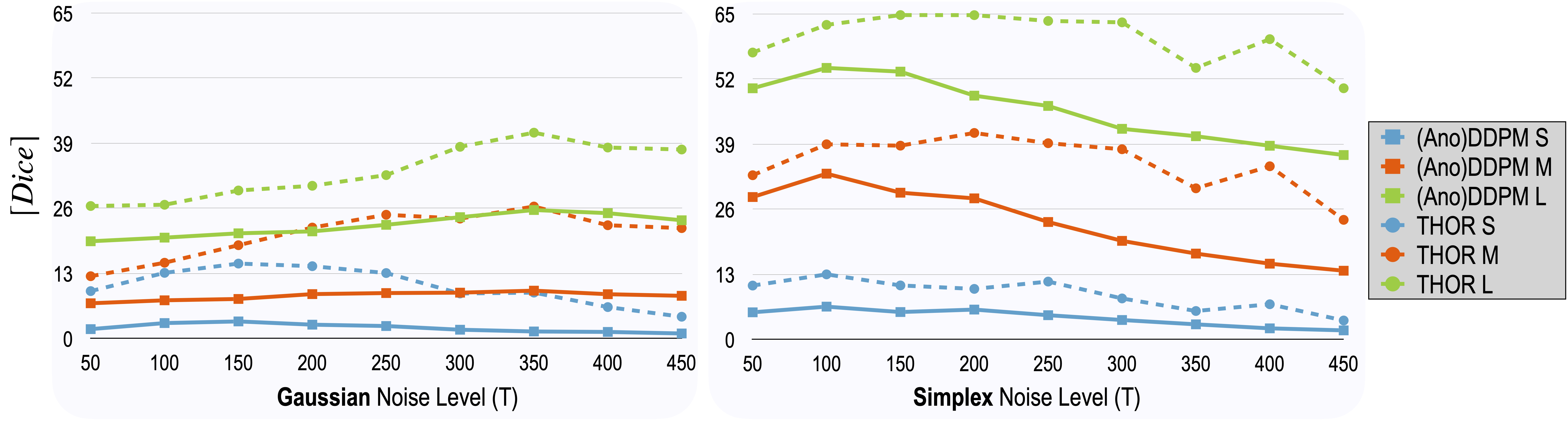} 
    \caption{\textbf{Noise Level Ablation.} \textit{THOR} outperforms the diffusion counterparts under both Gaussian and Simplex noise types across different noise levels $T$. \label{fig::noise_ablations}}
\end{figure}

\noindent\textbf{Sensitivity analysis of noise levels $T$} is shown in Fig.~\ref{fig::noise_ablations}. Increasing noise levels in the Gaussian setting enhances the detection of larger lesions, highlighting the role of higher noise in their effective masking. In contrast, performance declines with elevated Simplex noise levels, likely due to a self-supervision effect where anomalies mimicking the coarse noise pattern are identified and eliminated. While it achieves improved Dice scores for stroke segmentation, caution is advised. Its specificity to the coarse noise pattern may limit its effectiveness in broader anomaly detection scenarios by potentially overlooking anomalies that do not match this pattern. \textit{THOR} excels across different noise intensities, showcasing particular robustness at higher levels. This robustness minimizes the need for finely tuned noise adjustments for specific applications or anomaly sizes, underscoring \textit{THOR}'s adaptability and efficacy in anomaly detection tasks.


\subsection{Anomaly Localization in Pediatric Wrist X-rays \label{sec::wxr}}
Bone fractures are notably prevalent in children, with their detection being a critical step in ensuring timely medical intervention. \\

\noindent\textbf{Dataset.} We utilize the comprehensive GRAZPEDWRI-DX dataset~\cite{nagy2022pediatric}, encompassing 10,643 X-rays of pediatric wrist injuries from 6,091 individual patients. 
It includes a wide array of anomalies annotated with bounding boxes by certified pediatric radiologists. This includes bone anomalies (BA), foreign bodies (FB), fractures (Frac.), the presence of metal implants, periosteal reactions (PR), and soft tissue conditions (Soft). We report the recall and F1 scores.\\

\begin{figure}[t!]
    \centering
    \includegraphics[width=0.999\textwidth]{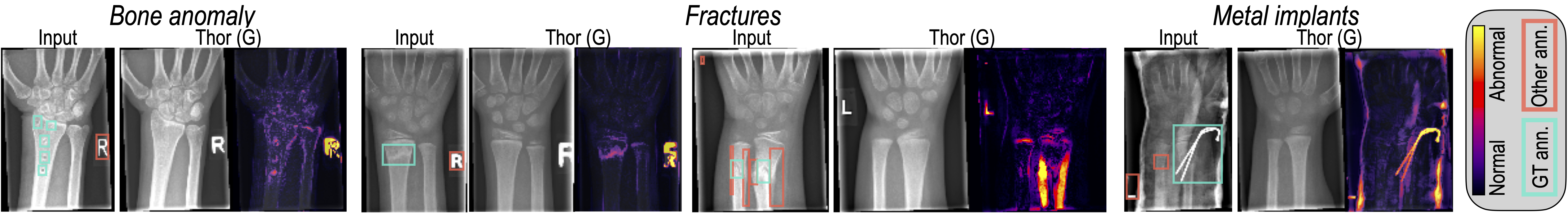} 
    \caption{Anomaly detection in pediatric wrist X-rays processed by \textit{THOR} using Gaussian noise. False positives arise from unannotated non-pathological changes like  unnatural bone positions following fractures or the presence of casts.}
    \label{fig::quali_wxr}
\end{figure}

\noindent\textbf{Results.} Tab.~\ref{tab::wrist} and Fig.~\ref{fig::quali_wxr} present both quantitative and qualitative outcomes. For this experiment, we concentrated on Gaussian noise, recognizing from prior analysis that Simplex noise lacks versatility for widespread anomaly detection applications. It tends to replicate anomalies dissimilar to its coarse patterns, such as bone anomalies, foreign bodies, and metal implants, as detailed in the supplementary materials. \textit{THOR} outperforms SOTA diffusion models, considerably improving the number of anomalies detected by up to 65\% in case of fractures and achieving almost perfect recall in detecting metal implants. 
\begin{table*}[t!]
    \centering 
    \caption{Anomaly detection and localization results in pediatric wrist X-rays.\label{tab::wrist}}
    \centering
    \setlength{\tabcolsep}{1pt}
        \begin{adjustbox}{width=1\linewidth,center} 
            \begin{tabular}{l  l| c c | c c | c c | c c | cc | cc }
                \toprule	    
        
                \multirow{2}{*}{\rotatebox{90}{Noise}} & \multirow{2}{*}{Method} & \multicolumn{2}{c|}{BA}  &  \multicolumn{2}{c|}{FB} & \multicolumn{2}{c|}{Frac.} & \multicolumn{2}{c|}{Metal} & \multicolumn{2}{c|}{Pr.} & \multicolumn{2}{c}{Soft}\\
                & & Recall  & F1 & Recall  & F1  & Recall  & F1  & Recall  & F1 & Recall & F1  & Recall  & F1 \\ \midrule
                \multirow{3}{*}{\rotatebox{90}{Gauss}} & \cellcolor{gray!10} THOR (ours) & \cellcolor{gray!10} {\boldmath $83.33$} & \cellcolor{gray!10} {23.76} & \cellcolor{gray!10} {\boldmath $75.00$} & \cellcolor{gray!10} 25.00 & \cellcolor{gray!10} {\boldmath $75.39$} & \cellcolor{gray!10} {\boldmath $16.46$} & \cellcolor{gray!10} {\boldmath $99.76$} & \cellcolor{gray!10} {\boldmath $73.76$} & \cellcolor{gray!10} {\boldmath $76.42$}& \cellcolor{gray!10} 16.64 & \cellcolor{gray!10} 26.32 &\cellcolor{gray!10}  10.77\\
                & DDPM~\cite{ho2020denoising} & 32.22 & 6.35 & {\boldmath $75.00$} & 29.83 & 28.53 & 5.10 & 86.47 & 39.66 & 52.25 & 9.79 & 23.68 & 8.89\\ 
                & AutoDDPM~\cite{bercea2023mask} & 63.89 & {\boldmath $23.93$} & {\boldmath $75.00$} & {\boldmath $58.33$} & 45.56 & 15.84 & 95.89 & 72.05 & 62.29 & {\boldmath $29.00$} &  {\boldmath $31.58$} & {\boldmath $16.45$} \\ 
         	    \bottomrule
            \end{tabular}
        \end{adjustbox}
\end{table*}

\section{Discussions and Conclusion}
This paper introduces \textit{THOR}, a diffusion-based framework for anomaly detection in medical imaging, which incorporates a novel harmonization process to enhance the denoising and restoration, thereby improving segmentation accuracy. We rigorously tested \textit{THOR} in two challenging scenarios: detecting strokes in brain MRIs and identifying pediatric wrist injuries in X-rays. Our results show that \textit{THOR} considerably outperforms existing diffusion methods.

However, unsupervised anomaly detection still faces challenges such of false positives due to unannotated non-pathological changes shown in Fig.~\ref{fig::qualitative} and Fig.~\ref{fig::quali_wxr}. These are correctly identified as anomalies, but not annotated by the radiologists. Furthermore, some conditions like soft tissue anomalies are subtle and difficult to spot on small resolutions. Additionally, we discovered that Simplex noise exhibits a self-supervision effect, introducing a bias in the anticipated anomaly distribution, warranting cautious use in wide-ranging anomaly detection tasks.

Our future efforts will focus on overcoming these obstacles to improve the diagnostic precision and broaden the utility of unsupervised anomaly detection across a diverse array of anomalies, organs, and imaging modalities.
\bibliographystyle{splncs04_unsort}

\bibliography{main}
\section{Appendix}
\begin{figure}[b!]
    \centering
    \includegraphics[width=0.990\textwidth]{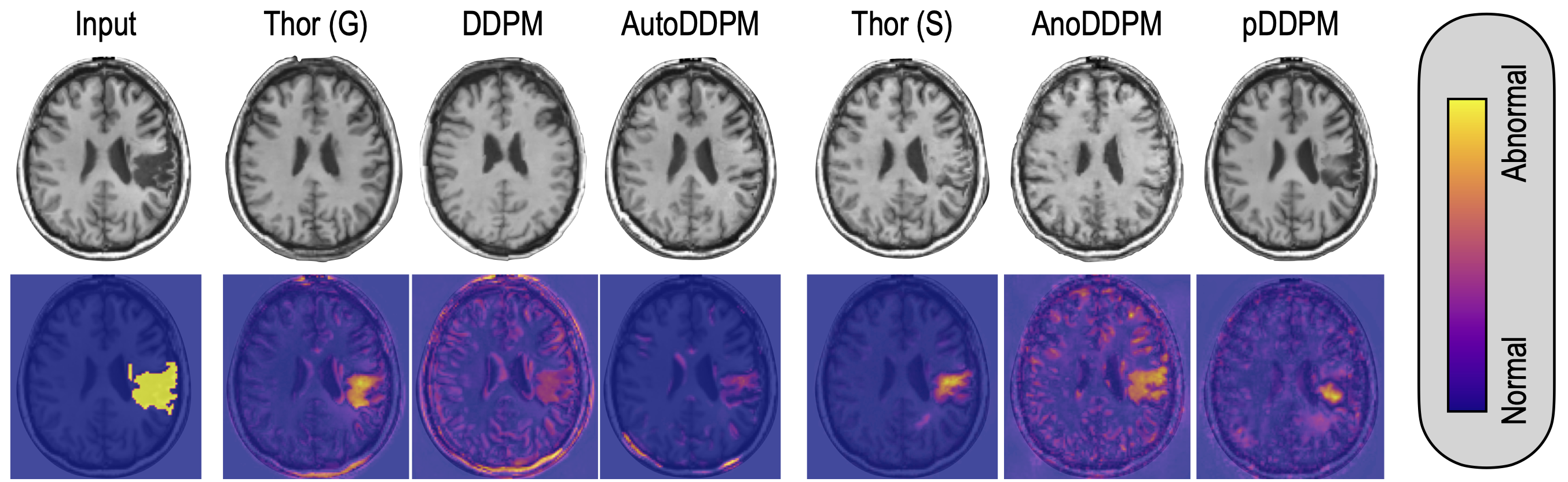} 
    \caption{Qualitative assessment of different diffusion-based models in Brain MRI. \textit{THOR} refines the performance of both DDPM (Gaussian) and AnoDDPM (Simplex), resulting in more accurate reconstructions and enhanced segmentations.}
    \label{fig::quali_all}
\end{figure}
\begin{figure}[tb!]
    \centering
    \includegraphics[width=0.999\textwidth]{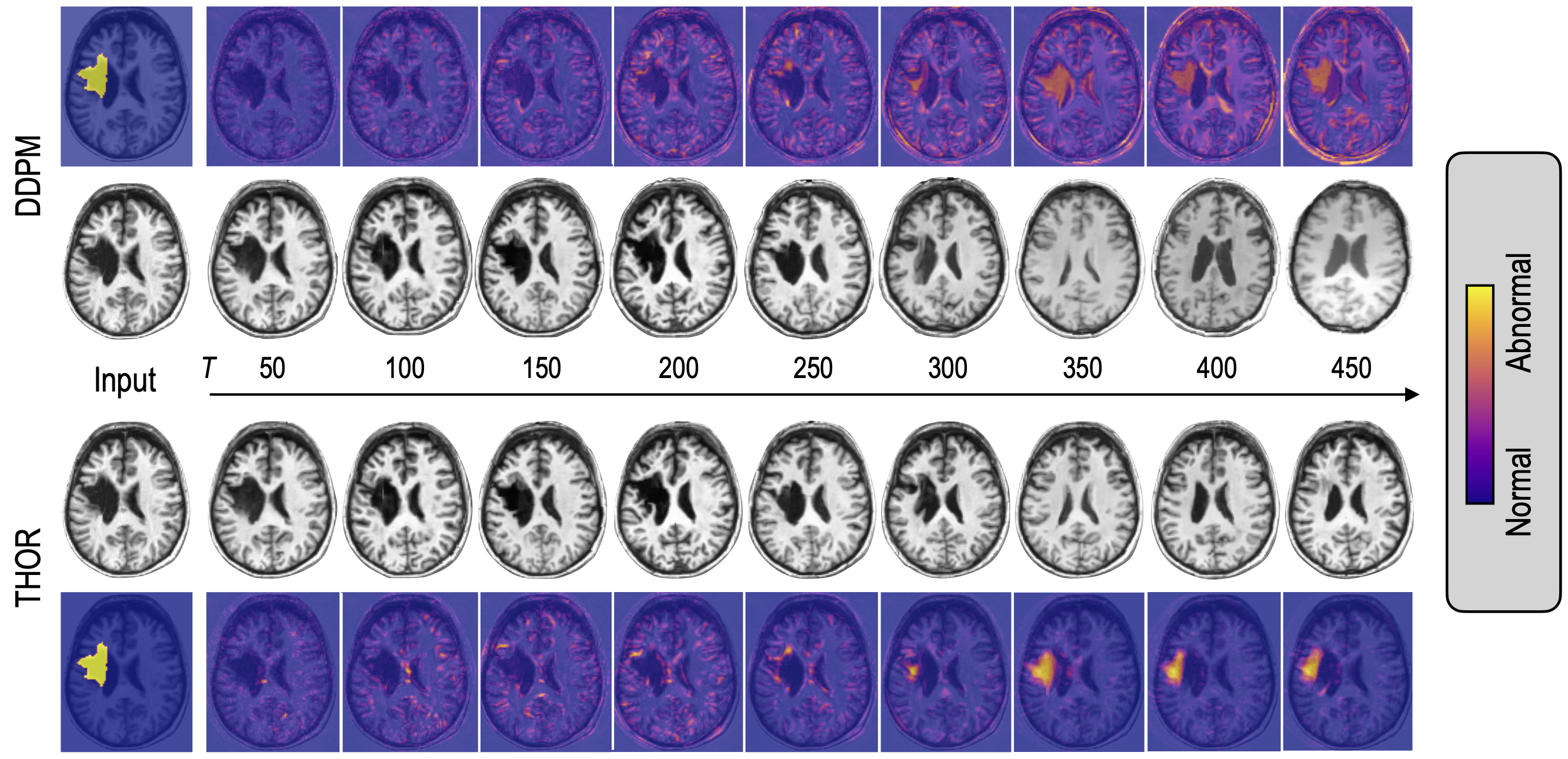} 
    \caption{Comparison of noise scales T for DDPM and \textit{THOR} using Gaussian noise reveals that low noise levels ($\leq 300$) retain the anomaly's structure, leading to missed detections. At noise levels $> 300$, DDPM diverges significantly from the original image, affecting even pathology-free areas. Conversely, \textit{THOR} applies temporal harmonization to generate outputs more closely aligned with the input, yet within the healthy spectrum, across all noise levels $> 300$.}
    \label{fig::noise_quali}
\end{figure}
\begin{figure}[tb!]
    \centering
    \includegraphics[width=0.999\textwidth]{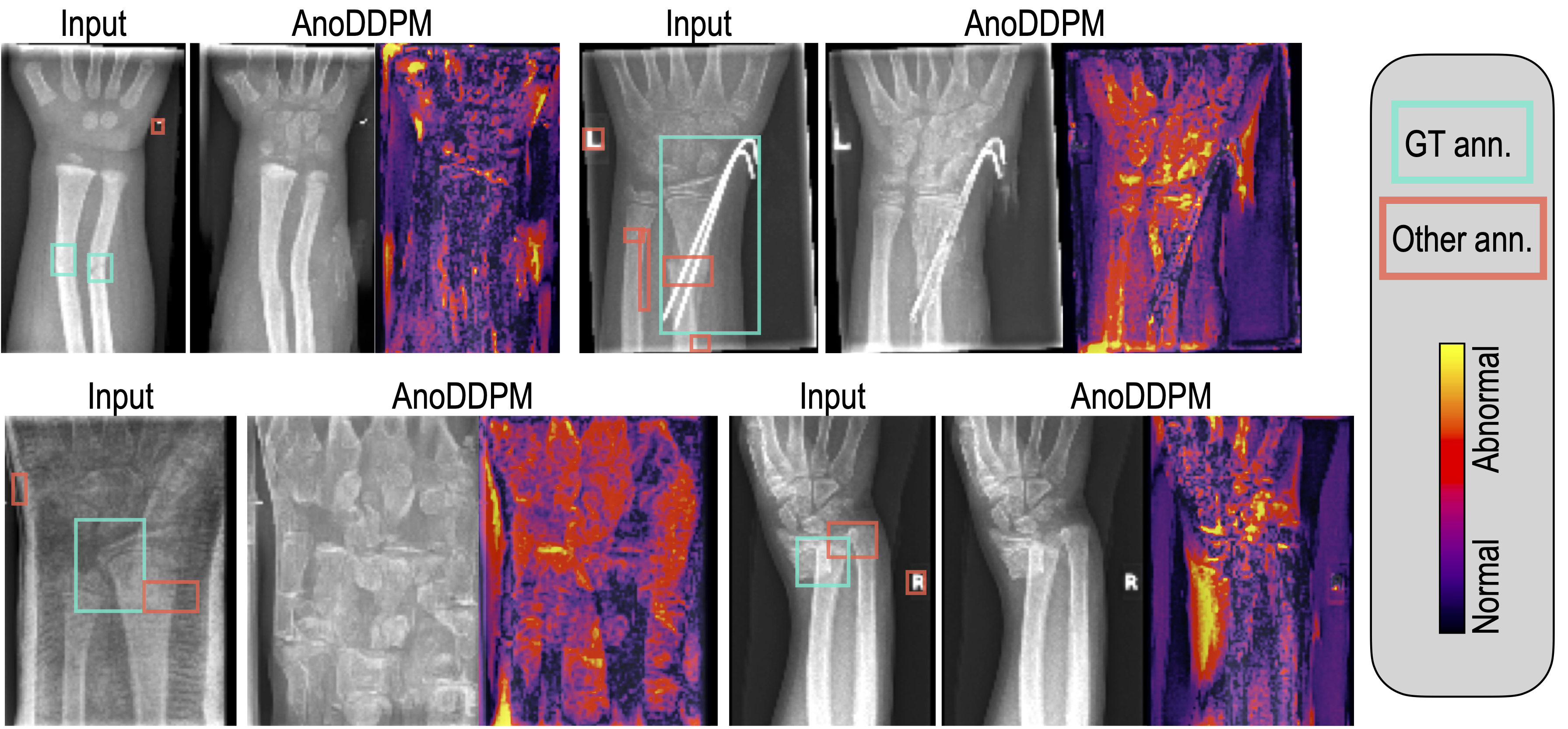} 
    \caption{AnoDDPM's performance with Simplex noise in detecting anomalies in wrist X-ray images demonstrates that the diffusion process reconstructs fractures, unnatural bone positions, and metal implants. Similar to observations in brain MRI experiments, Simplex noise struggles to accurately learn the healthy anatomy, instead, it aims at eliminating structures akin to its coarse noise patterns. Although literature cites its successful application in tumor segmentation, and our work confirms enhanced performance in stroke lesion segmentation, Simplex noise proves unsuitable for general anomaly detection tasks.}
    \label{fig::anno_wxr}
\end{figure}
\end{document}